\documentclass[letter]{aa} 

\usepackage{epsfig}	
\usepackage{graphicx,color}	
\usepackage{amssymb}		
\usepackage{url}		
\usepackage{amsmath}		
\usepackage{rotating}			
\usepackage{float}			
\usepackage{textcomp}		
\usepackage{epstopdf}
\usepackage{dcolumn}
\usepackage{times}
\usepackage{tabularx}
\usepackage{hyperref}
\hypersetup{
    colorlinks,
    citecolor=blue,
    filecolor=blue,
    linkcolor=blue,
    urlcolor=blue,
    menucolor=black}
\usepackage{soul} 
\usepackage[english]{babel}
\usepackage{booktabs}

\newcommand{\HRIEUV}{HRI$_{\rm{EUV}}$}

\begin{document}


\title{Propagating brightenings in small loop-like structures in the quiet Sun corona: Observations from Solar Orbiter/EUI}

\author{Sudip Mandal\inst{1}, Hardi Peter\inst{1}, Lakshmi Pradeep Chitta\inst{1}, Sami K.~Solanki\inst{1,2}, Regina Aznar Cuadrado\inst{1}, Luca Teriaca\inst{1}, Udo~Sch\"{u}hle\inst{1}, David~Berghmans\inst{3}
\and
 Fr\`{e}d\`{e}ric~Auch\`{e}re\inst{4}}

   \institute{Max Planck Institute for Solar System Research, Justus-von-Liebig-Weg 3, 37077, G{\"o}ttingen, Germany \\
              \email{smandal.solar@gmail.com}
   \and
             School of Space Research, Kyung Hee University, Yongin, Gyeonggi 446-701, Republic of Korea
   \and
             Royal Observatory of Belgium, Ringlaan -3- Av. Circulaire, 1180 Brussels, Belgium
   \and
             Institut d'Astrophysique Spatiale, CNRS, Univ. Paris-Sud, Universit\'{e} Paris-Saclay, B\^{a}t. 121, 91405 Orsay, France
}

\abstract{
Brightenings observed in the solar extreme-ultraviolet (EUV) images are generally interpreted as signatures of micro- or nanoflares occurring at the transition region or coronal temperatures. Recent observations with the Extreme Ultraviolet Imager (EUI) on board Solar Orbiter have revealed the smallest of such brightenings (termed campfires) in the quiet-Sun corona. Analyzing EUI 174~{\AA} data at a resolution of about 400 km on the Sun with a cadence of 5\,s from 30-May-2020, we report here a number of cases where these campfires exhibit propagating signatures along their apparent small (3-5 Mm) loop-like structures. Measured propagation speeds are generally between 25~km~s$^{-1}$ and 60~km~s$^{-1}$. These apparent motions would be slower than the local sound speed if the loop plasma is assumed to be at a million Kelvin. Furthermore, these brightenings exhibit non-trivial propagation characteristics such as bifurcation, merging, reflection and repeated plasma ejections. We suggest that these features are manifestations of the internal dynamics of these small-scale magnetic structures and could provide important insights into the dynamic response ($\sim$40~s) of the loop plasma to the heating events as well as into the locations of the heating events themselves.}

   \titlerunning{Propagating brightenings in small loop-like structures}
   \authorrunning{Sudip Mandal et al.}
   \maketitle

\section{Introduction}

Localized brightenings are commonly detected using ultraviolet (UV) or extreme ultraviolet (EUV) imaging and spectroscopic diagnostics of the solar chromosphere through the corona. These brightenings are generally compact, dynamic and are thought to be driven by magnetic reconnection. Depending upon their imaging and spectroscopic signatures (including energetics), these bright features are often referred to in the literature as bright points \citep{2003A&A...398..775M}, explosive events \citep{1983ApJ...272..329B,1989SoPh..123...41D,1997Natur.386..811I}, blinkers \citep{1997SoPh..175..467H}, or UV bursts \citep{2014Sci...346C.315P}. For a comprehensive review on this subject see \citet{2018SSRv..214..120Y}. Multi-instrument observations further reveal that these bright features appear everywhere on the solar disc, i.e from quiet-Sun areas to active region neighborhoods \citep{1997ApJ...488..499K,1998A&A...336.1039B,1999SoPh..186..207B,2000ApJ...535.1027A,2019ApJ...887...56T}. Being nearly ubiquitous, these brightenings were initially thought to be potential candidates for coronal heating \citep{1991SoPh..133..357H}. However, further investigations point out that one would need significantly more ($\sim$100 times) of these events in order to account for the necessary energy budget \citep{2000ApJ...535.1047A,2021A&A...647A.159C}.

Recent observations from the Extreme Ultraviolet Imager \citep[EUI;][]{2020A&A...642A...8R} on board Solar Orbiter \citep[][]{2020A&A...642A...1M}, have uncovered localized brightenings (termed as campfires) down to sizes as small as 0.08~Mm$^2$ \citep{2021arXiv210403382B}. These are the smallest such events yet observed in the quiet-Sun corona. Although most of these campfire events observed with EUI are also detected with the Atmospheric Imaging Assembly onboard the Solar Dynamics Observatory \cite[AIA/SDO;][]{2012SoPh..275...17L}, their AIA counterparts appear to be rather faint and fuzzy. This is mostly due to the coarser spatial and temporal resolutions of AIA. In an effort to explain the physical origin of these brightenings, \citet{2021arXiv210403382B} proposed a scenario in which the apex of a low lying\footnote{Using stereoscopic techniques, \citet{2021arXiv210902169Z} estimated the heights of the campfires to be between 1000 km and 5000 km above the photosphere.} coronal (or transition region) loop gets heated due to (component) magnetic reconnection and subsequently, the heated plasma appears as a localized brightening. That interpretation is substantiated by \citet{2021arXiv210410940C} who find such campfire-like events in a 3D radiation magnetohydrodynamic (MHD) simulation. These authors have also demonstrated that there could be different possible magnetic configurations in which these campfires are triggered (e.g., forking field lines, crossing field lines).

Reports of subarcsec brightenings using high resolution data from past missions such as IRIS \citep{2014SoPh..289.2733D} and Hi-C \citep{2014SoPh..289.4393K,2019SoPh..294..174R}, have mostly been limited to regions that are either close to or within an active region \citep{2013ApJ...771...21W,2014ApJ...790L..29T,2014Sci...346C.315P,2016ApJ...822...35A,2019ApJ...887...56T}.  \citet{2014ApJ...790L..29T} found subarcsec brightenings in IRIS data within sunspot penumbrae that have speeds between 10--40~km~s$^{-1}$. On the other hand, analyzing Hi-C data, \citet{2013ApJ...771...21W} reported localized plasma flows at high speeds (between 90--290~km~s$^{-1}$) within small loops that are embedded inside a moss region. Recently, a statistical study of EUV bursts in quiet-Sun regions has been presented in \citet{2021A&A...647A.159C}. The detection of these burst-like features was primarily constrained by the coarser spatial resolution of AIA data (about 900\,km) used in that study. Furthermore, this limitation also meant that the internal dynamics of such EUV bursts could not be studied with AIA. 

\begin{figure*}[!htb]
\centering
\includegraphics[width=0.80\textwidth,clip,trim=0cm 0cm 0cm 0cm]{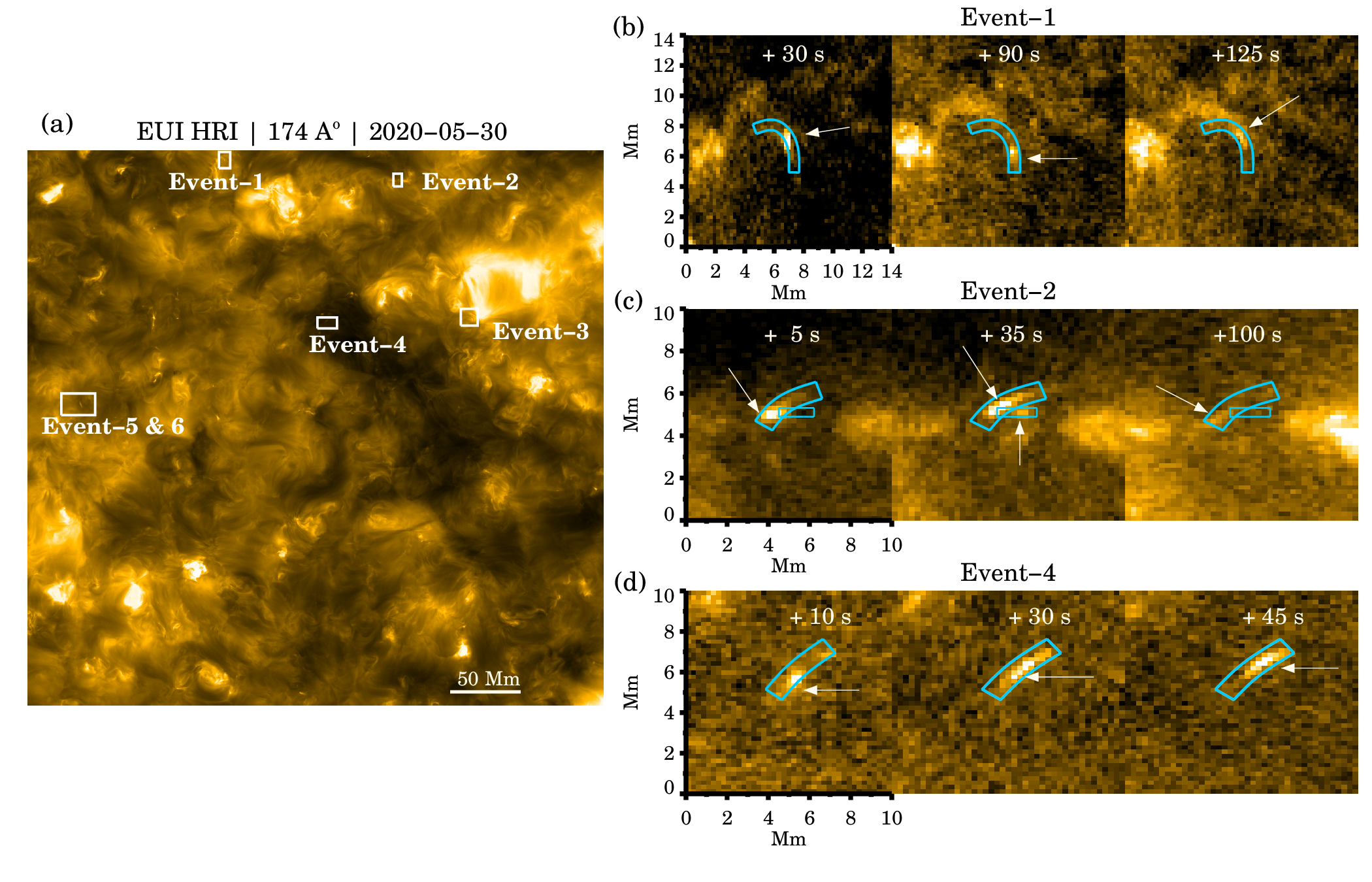}
\caption{Internal dynamics in small loop-like features. {\it Panel-a} shows the time averaged image of the {\HRIEUV} observation from 30-May-2020. White boxes outline the locations of the six events studied in this paper. {\it Panels-b,d} present snapshots from Events 1, 2 and 4, respectively. In each of these snapshots, the blue curved box marks the extent of the artificial slit that is used to derive the space-time map, whereas the arrow points to the instantaneous position of the localized brightening in that frame. The elapsed time for each frame is given in seconds as measured from 14:54:00 UT, i.e., the start time of the analyzed observation.}
\label{fig1}
\end{figure*}
In this letter, we analyze the EUI data taken during the commissioning phase of the Solar Orbiter mission \cite[][]{2020A&A...642A...1M} and present a number of localized brightening events in the quiet-Sun region that exhibit systematic propagation signatures along their host loop-like structures. Thanks to its higher spatial resolution (about 400\,km) as well as finer image cadence (of 5~s) compared to AIA data, the EUI imaging data facilitate us to study the dynamical nature of these brightenings in greater detail. We describe the data in Section 2, whereas Section 3 outlines the results. Finally we conclude by summarizing our results in Section 4.

\section{Data}\label{sec_data}
We use  EUV imaging data\footnote{We use the level\,2 data (L2) which can be accessed via \url{https://wwwbis.sidc.be/EUI/data/releases/202107_release_3.0/}. Here also information on the data processing can be found in the release notes. DOI: \url{https://doi.org/10.24414/k1xz-ae04}.} from the High Resolution Imager (HRI) of the Extreme Ultravoilet Imager \cite[EUI;][]{2020A&A...642A...8R}, onboard Solar Orbiter. These images were taken on 30-May-2020 with the {\HRIEUV} telescope that images a wavelength band around 174~{\AA} which captures the dynamics of the solar plasma with temperatures between ${\log}\,T$\,[K]${=}$5.3 and 6.3 (i.e., transition region to cooler corona) with a peak near 1\,MK. We align the individual images using the ssw routine {\rm fg\_rigidalign.pro} which removes the effect of jitters in the data by means of cross-correlations. The image cadence is 5~s and a total of 50 frames were recorded. The pixel scale, in each direction, is 0.492\arcsec. At the time of this observation, Solar Orbiter was located at a distance of about 0.556 astronomical units from the Sun. Hence, each pixel in these images corresponds to a distance of 198 km on the solar surface at disc center. Further details about this observational campaign can be found in \citet{2021arXiv210403382B}.

\section{Results}\label{sec_result}

\subsection{General characteristics}\label{sec_gen_char}

Fig.~\ref{fig1}a shows the full 406$\times$406 $\rm{Mm^2}$ field of view (FOV) of the analyzed observation, which primarily encompasses a quiet-Sun region near the disc center (as seen from the vantage point of Solar Orbiter). The overplotted white boxes outline the locations of the six events that we analyze in this paper. Except one case (Event-3), all other events are located within relatively quiescent regions in the FOV. A feature that is common to all these events is the systematic propagation of a localized brightening along the loop-like host structure (see Figs.\,\ref{fig1}(b)--(d) and online animations). These propagating brightenings are mostly between 5 to 20 pixels (i.e., between 0.2 to 0.8 Mm$^2$) in size and their appearance is comparable to the moderate sized campfires reported in \citet{2021arXiv210403382B}. We also note that the structures hosting these brightenings appear to be small loop-like configurations with footpoint separations of 2 to 4 Mm. 

In order to highlight the overall temporal characteristics of these events, snapshots from three of these events are presented in Fig.~\ref{fig1}b (Event-1), Fig.~\ref{fig1}c (Event-2) and Fig.~\ref{fig1}d (Event-4). As can be seen through these plots, the analyzed compact brightenings appear to propagate, although the propagation characteristics in each of these events are distinctly different. For example, the brightening in Event-1 first appears near the middle of the structure, whereas in the other two examples, it appears close to the footpoints. The other aspect to note here is the spatial extent (and shape) of these features. In Event-1 the brightening is rather compact (dot-like) whereas in Event-4 it is quite elongated (loop-like). Similar shape distinctions were also reported in \citet{2019ApJ...887...56T,2021arXiv211006846P}.

\begin{figure*}[!htb]
\centering
\includegraphics[width=0.97\textwidth,clip,trim=0cm 0cm 0cm 0cm]{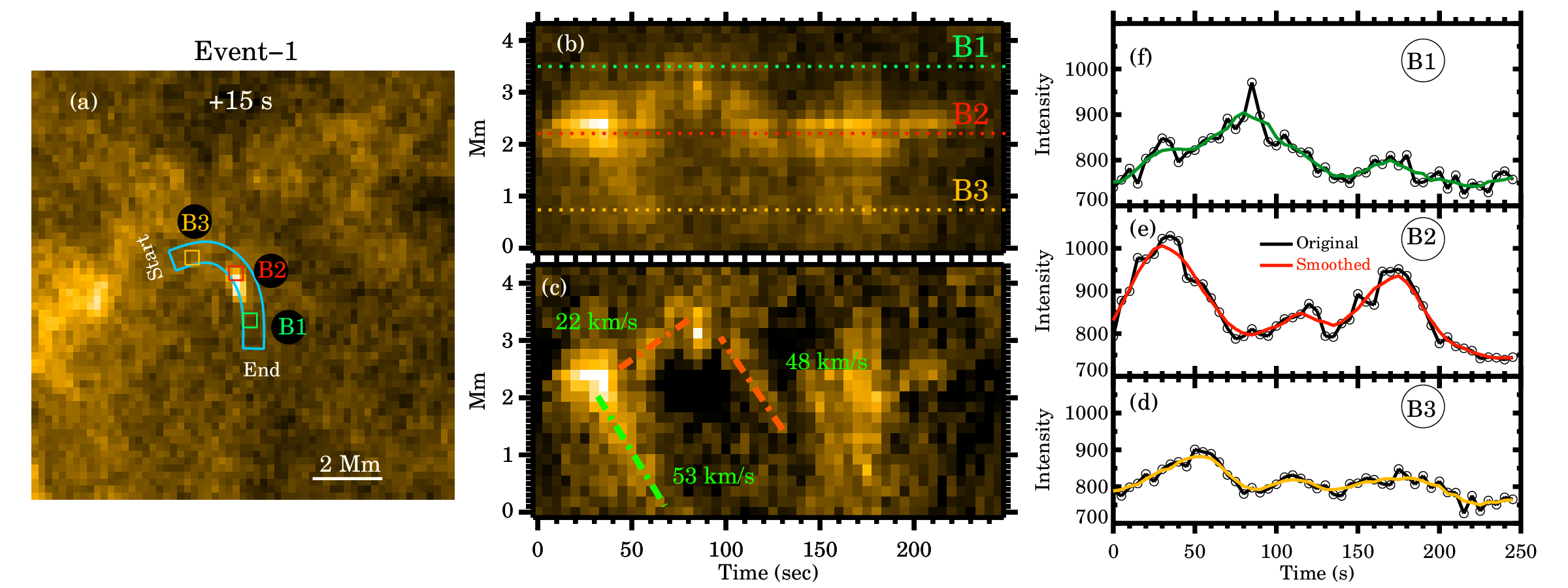}
\caption{Bifurcation scenario 1 (Event-1). {\it Panel-a} shows the context image for this event along with the artificial slit (the blue curved box) that we use to generate the space-time (X-T) map ({\it Panel-b}). The background subtracted version of this map is shown in {\it Panel-c}. The label `start' (`end') marked on {\it Panel-a}, refers to the origin (end point) of the y-axis of the X-T map. The inclined green and orange dashed-dotted lines in {\it Panel-c} represent the slopes of the individual ridges used for speed estimations. In-fact, the orange dashed-dotted lines encapsulate the inverse-v shape that indicates a reflection at the loop footpoint. {\it Panels-d,e,f} illustrate the light curves that we extract from the three square boxes, $\rm{B1}$, $\rm{B2}$ and, $\rm{B3}$ as shown in {\it Panel-a}. Center points of these boxes are highlighted as horizontal lines in {\it Panel-b}. For more details, see Sect.\ref{sec_ev1}. (An animated version of this figure is available online)
}
\label{fig2}
\end{figure*}

To explore more about these aspects, we present in-depth descriptions of each of these events in the following sections. In the main paper, we present four of these events wheres the remaining two are discussed in the Appendix.

\subsection{Event 1: Bifurcation scenario 1 }\label{sec_ev1}

\begin{figure*}[!htb]
\centering
\includegraphics[width=0.95\textwidth,clip,trim=1cm 18cm 0cm 0cm]{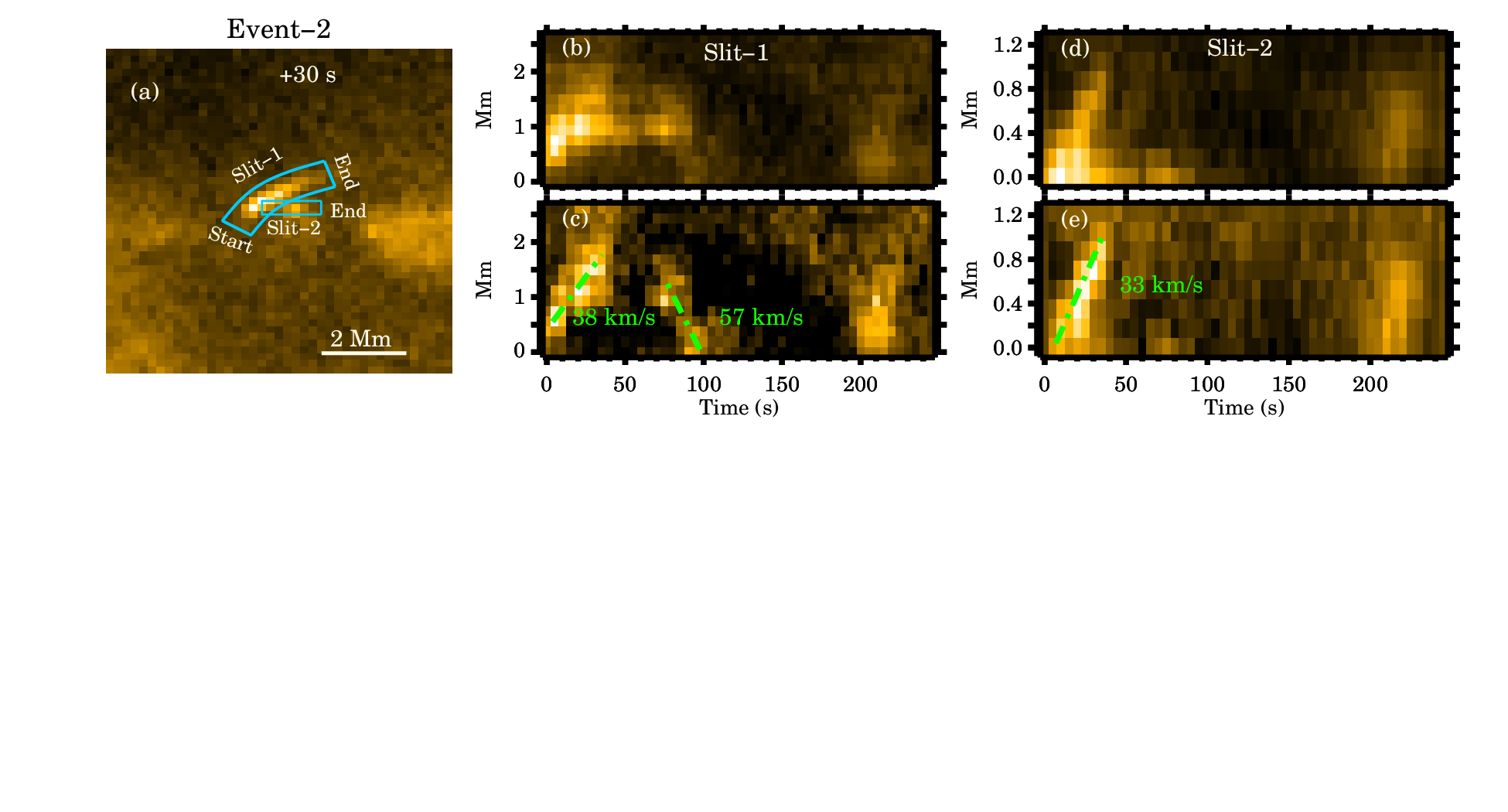}
\caption{Overview of bifurcation scenario 2 (Event-2). The layout is similar as in Fig.~\ref{fig2}a-c. {\it Panel-a} shows the context image. The blue curved boxes shown in this panel mark the locations of the artificial slits (Slit-1 and Slit-2). {\it Panels-b,c} shows the original and the background subtracted space-time maps for Slit-1. The same but for Slit-2 are shown in {\it panels-d,e}. The green dashed lines indicate the slopes of the ridges used for speed estimations. See Sect.~\ref{sec_ev2} for details. An animated version of this figure is available online.
}
\label{fig3}
\end{figure*}

This particular event occurs close to the upper edge of the observation's FOV (see Fig~\ref{fig1}a). The host, in this case, appears to be a small loop whose plane-of-sky length is approx. 4 Mm. To capture the propagation signatures, we first construct a space-time (X-T) map by placing an artificial slit that traces the propagation path of the observed brightening. The blue curved box in Fig.~\ref{fig2}a highlights this artificial slit (The X coordinate runs from the  top-left end of the curved slit to its bottom). The corresponding X-T map is shown in Fig~\ref{fig2}b. Additionally, to enhance the contrast of this map, we detrend (and normalize) the intensity time series at every spatial location using a 150~s (30 frames) running average. Fig~\ref{fig1}c shows this enhanced\footnote{The contrast enhancement procedure does preserve all the features of the original X-T map.} X-T map. From this map, we find that the localized brightening first appears near the middle of the loop at $t$=10~s. At $t$=35~s it bifurcates into two smaller brightenings. These two features then start moving in opposite directions (i.e. towards the apparent footpoints of the host structure). 

In the X-T map, this diverging propagation is seen as two oppositely slanted ridges between $t$=35~s and $t$=80~s. The propagation speeds are measured\footnote{These estimations are based on the slope of a straight line that is representative of the slope of a given X-T map ridge, as inferred visually. Furthermore, all speeds we refer here are the projected or the plane-of-sky speeds.} as 22~km~s$^{-1}$ and 53~km~s$^{-1}$. Considering that the loop plasma is at million Kelvin (which is reasonable given that the 174~{\AA} response function peaks at $\approx$1~MK), these {bf speed} are well below the local sound speeds (which is $\approx$150~km~s$^{-1}$). One of the propagating features (the one between B2 and B1;  Fig.~\ref{fig2}a) moves from one end of its path to the other after which it starts returning along its original path (as highlight with white arrows in Fig.~\ref{fig1}b). This returning feature moves with a speed of 48~km~s$^{-1}$ towards the apparent loop apex (i.e. toward B2). It is this back and forth motion which produces the inverted v-shape (reflection-like) signature in the X-T map\footnote{The ridge pattern seen in Fig.~\ref{fig2}b and c between X=2 to 3.5 Mm and T=40 to 140 s is being referred to as the inverted v-shape (reflection-like) signature.  This shape is outlined by the two orange lines in Fig.~\ref{fig2}c.}. Interestingly, no such return feature is observed for the other brightening that moved towards the other footpoint (i.e. B3). Later at $t$=170~s, a new brightening appears near the loop apex, which covers a significant portion of the loop (evident through the vertical extent of the feature in the X-T map). In comparison, the initial brightening around $t$=35~s is more localized before it exhibits propagation. Moreover, the second brightening (at $t$=170~s) showed no clear sign of propagation (as deemed visually) during its entire lifetime of $\approx$80~s.

In order to further investigate the temporal evolution of the observed phenomenon, we now analyze light curves from three different locations within this loop-like structure. Three boxes, each of 3$\times$3 pixels in size, mark the locations (Fig.~\ref{fig2}a) where we extract the light curves: $\rm{B1}$ near the right footpoint, $\rm{B2}$ at the apex, and $\rm{B3}$ near the other footpoint. The extracted light curves are presented in Panels~\ref{fig2}d-e. There are a couple of interesting features to highlight. Light curves from the two opposite ends of the loop, i.e. from $\rm{B1}$ and $\rm{B3}$, show rising trends well before the propagating brightenings from $\rm{B2}$ actually reached there. This hints towards the following scenario: although the main reconnection takes place near the apparent loop apex, the footpoints get excited simultaneously, which is similar to near-simultaneous footpoint brightenings observed in active region core loops \citep[see][]{2020A&A...644A.130C}. It is, however, also possible that (undetected) fainter propagating brightenings from $\rm{B2}$ have actually excited the footpoints $\rm{B1}$ and $\rm{B3}$. In this scenario, the observed delay between the first peak in $\rm{B1}$ ($\rm{B3}$) and that in $\rm{B2}$ occurs due to the finite propagation time that the brightening at $\rm{B2}$ takes to reach to that footpoint. Additionally we note that, at the time of the second peak in $\rm{B2}$ (associated with reflection from B1), we also find a weak second peak (at $t$=115~s) in $\rm{B3}$. This could mean that there exists a faint `reflected' brightening from this footpoint, too. However, such a feature is rather unrecognizable in the original X-T map (Fig.~\ref{fig2}b) as well as in the event movie. The last peak (at $t$=170~s) in all three light curves is basically due to the new extended brightening that appears again near the middle of the loop.


\subsection{Event 2: Bifurcation scenario 2}\label{sec_ev2}

In Event-2 (Fig~\ref{fig3}a), we find the brightening to first appear near the left footpoint of its host loop. The plane of sky extent of this loop is approx 2 Mm. Now, as soon as the brightening starts moving towards the right footpoint, a small bright feature detaches itself from this initial brightening and propagates away from the loop. Therefore, this is similar to the `bifurcation' scenario we found in Event-1. With time, this small bifurcated feature fades away rather quickly after traveling a short distance away for the loop, whereas the initial brightening could be traced to the other footpoint before it reflects back to its starting location. Slit-1 and 2 in Fig~\ref{fig3}a outline the propagation paths of the initial and the detached brightenings, respectively. The space-time maps created using these two slits are shown in Fig~\ref{fig3}b-e (and are created in the same way as described in Sect.~\ref{sec_ev1}). In both of these maps, we again find the propagation speeds to be sub-sonic (ranging between 33~km~s$^{-1}$ and 57~km~s$^{-1}$). Lastly, we also note that the bright feature seen at $t$=205~s in Fig~\ref{fig3}b (also in Fig~\ref{fig3}d) is due to an extended brightening that appears momentarily without clearly detectable propagation characteristics. 

\subsection{Event 3: Merger}\label{sec_ev3}

\begin{figure}[!htb]
\centering
\includegraphics[width=0.48\textwidth]{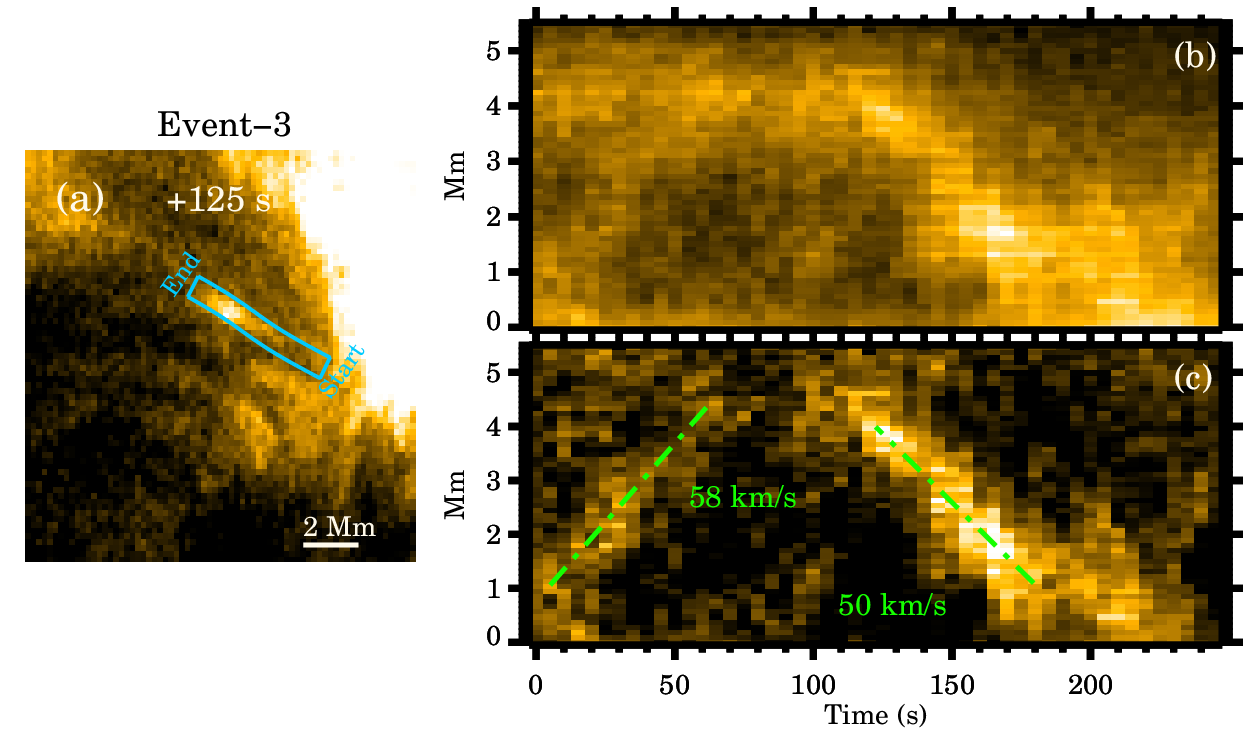}
\caption{Merger-type event (Event-3). The layout is the same as in Fig.~\ref{fig2}a-c. For details see Sect. \ref{sec_ev3}. An animated version of this figure is available online.}
\label{fig4}
\end{figure}

While Events 1 and 2 display apparent bifurcations of the intensity propagation, Event-3 (Fig~\ref{fig4}a) exhibits a different propagation signature: Here we find two separate brightenings to come together and become a single entity,  they merge. The initial two brightenings, one each at the two footpoints of the host structure, were already present on the first frame of this observation. Over time, the brightening at the right footpoint (labeled `start' in Fig.~\ref{fig4}a) moves towards the other footpoint (labeled `end'). Interestingly, throughout this time, the other brightening at the left footpoint remains stationary. In the X-T map (Fig~\ref{fig4}b), the stationary brightening appears as a horizontal bright ridge near the top of the map, wherein the slanted ridge, appearing between $t$=5~s and $t$=80~s, is due to the upward moving brightening. Subsequently, these two brightenings meet near the left footpoint and then the combined structure propagates back towards the right footpoint. As a result of this, we find a downward slanted bright streak in the X-T map between $t$=100~s and $t$=170~s. At this point the intensity of the return brightening is higher than that of the original outward directed one. Speeds of the outward and return propagation are measured as 58~km~s$^{-1}$ and 50~km s$^{-1}$, respectively (Fig.~\ref{fig4}c). The location of this event is not strictly within a quiet-Sun patch, but rather close to a region with enhanced emission.

\subsection{Event 4: Multiple ejections}\label{sec_ev4}

\begin{figure}[!htb]
\centering
\includegraphics[width=0.48\textwidth]{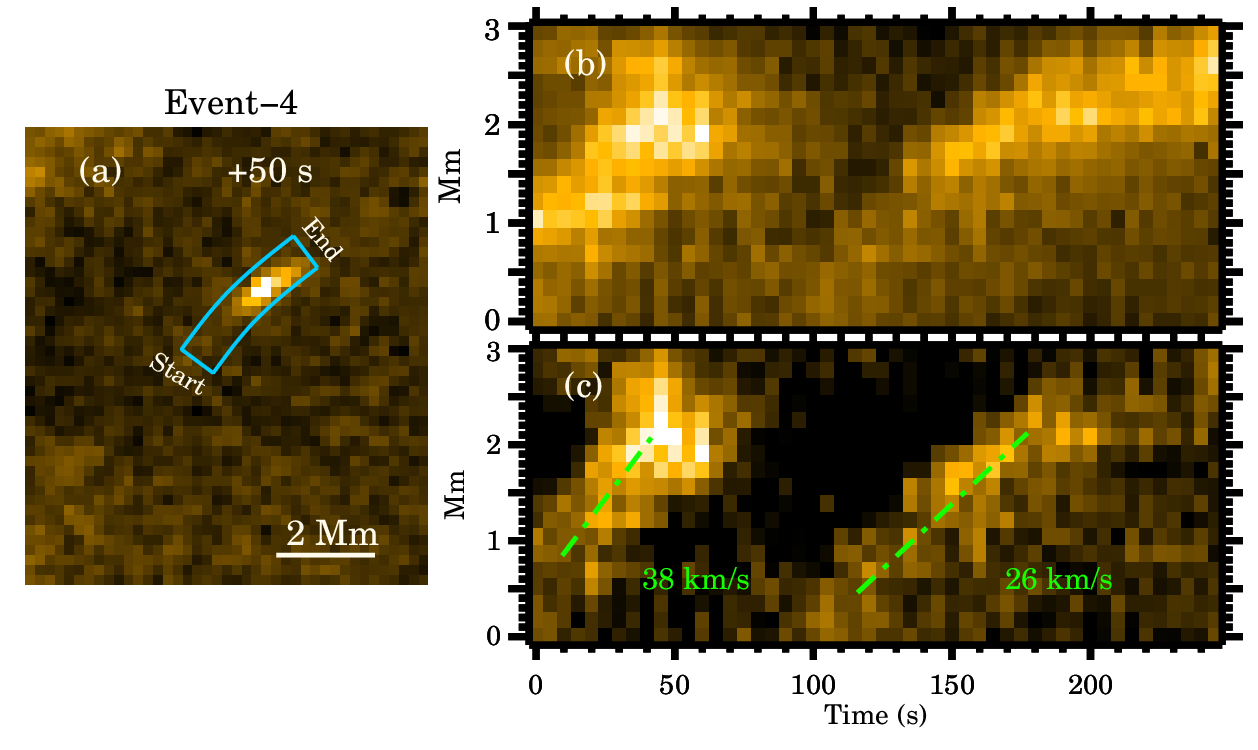}
\caption{Multiple-ejection-type event (Event-4). The layout is the same as in Fig.~\ref{fig2}a-c. For details see Sect. \ref{sec_ev4}. An animated version of this figure is available online.}
\label{fig5}
\end{figure}
Event-4 (Fig.~\ref{fig5}a) is somewhat different compared to the events we have discussed so far. In this case we observe two successive brightenings to propagate from one end of the structure to the other end (Fig.~\ref{fig5}b). These two ejections occur 50~s apart and their speeds are estimated as 38~km~s$^{-1}$ and 26~km~s$^{-1}$, respectively (Fig.~\ref{fig5}c). A distinct propagation characteristics is that the second ridge in the X-T map, becomes flatter at later times (around $t$=180~s). There is, however, no such clear trend in the first ridge, indicating that the second brightening might have reached further towards that footpoint and due to the geometry of a nearly vertical loop, the brightening appears to be decelerating. The lack of a seeming acceleration at the beginning of its journey suggests that either the brightening started higher up in the loop, or it started with a large initial speed and slowly decelerated. Furthermore, the intensity of the second brightening does increase gradually while it moves along the loop (Fig.~\ref{fig5}b).  This effect is rather weakly visible in the first brightening and can be attributed to the fact that the brightening already moved partway along the slit even before the time series began.
 
\section{Discussion and conclusion}\label{sec_conclusion}

In this letter, we present evidence of propagating brightenings in small loop-like structures as observed in timeseries of EUI high-resolution images in the 174\,{\AA} channel.

The identification of the shape of the loop we investigate, in particular the location of their apex and footpoints, are based solely on the propagation characteristics of the brightenings. For a more reliable identification one would need the underlying magnetic field information at a comparably high spatial resolution. While not being available for these data from the early mission phase, co-observations with the Polarimetric and Helioseismic Imager on board Solar Orbiter \citep[SO/PHI,][]{2020A&A...642A..11S}  will become available for future studies.

By analyzing multiple of the propagating coronal brightenings in campfire loop-like events we find that 
 
 \begin{enumerate}
 
\item
the initial brightening appears either near the middle (apex) of the loop or near to one of the loop footpoints,

\item
except for one case, the apparent propagation speeds lie between 25~km~s$^{-1}$ and 60~km~s$^{-1}$ which is below the local sound speed.

\item
these brightenings exhibit non trivial propagation characteristics such as bifurcation, back and forth motion, apparent merging of brightenings and repeated ejections.
 \end{enumerate}

These features are distinctly different from the bi- and unidirectional plasma jets that have recently been reported by \citet{2021arXiv210915106C}. Using an EUI data set of 2~s image cadence, \citet{2021arXiv210915106C} found compact jets at high speeds between 100~km~s$^{-1}$ and 150~km~s$^{-1}$. These are not only significantly faster than the propagating brightenings we report here, but they are also different in morphology and propagation characteristics, which points to a different driving mechanism.

As discussed earlier in Sect.~\ref{sec_gen_char}, the brightenings whose propagation characteristics we analyzed here are mostly campfire-like features \citep{2021arXiv210403382B}. Although campfires are the smallest such events yet observed in the quiet-Sun corona, it is not yet clear whether they are (physically) different from some of the already known brightening phenomena e.g., coronal bright points or the subarcsec transition region brightenings. To disentangle these scenarios, one would need to use simultaneous spectroscopic data \cite[e.g., from SPICE,][]{2020A&A...642A..14S} which are not available for the analyzed observations. Nevertheless, based on their rapid evolution, it is quite evident that these brightenings are likely products of magnetic reconnection. There could be different possible topologies originating from the underlying magnetic fields that are prone to reconnection \citep{2012ApJ...746...19Z,2015ApJ...813...86I,2017A&A...606A..46G,2017A&A...605A..49C,2019LRSP...16....2M}. Nonetheless, we can use the propagation characteristics of the observed brightenings to infer the underlying magnetic topology. For example, the bifurcation scenario we find in Event-1 could be explained by a reconnection between two crossing field lines. The bifurcation (and subsequent propagation) seen in Event-2 may possibly be a signature of component reconnection occurring in tangled field lines (e.g., similar to the one shown in Fig.4d of \citealp{2021arXiv210410940C}). Due to the lack of information about the underlying photospheric magnetic field at matching spatial resolution\footnote{Footpoints of these campfire loops are not resolved with SDO/HMI.} we cannot confirm or disprove these different scenarios at this point. Co-observations with SO/PHI that will be available in future observations will  provide more insights into this.

In two of our events (Event-1 and 3) we have found  brightenings that, after travelling from one end of the host (loop-like) structure to its other end, were reflected back. These cases appear similar to gentle chromospheric evaporation, in which non-thermal electrons from the site of a flare meet the chromospheric material at the footpoints of the loop they are propagating along and produce evaporated material upflow at a speed of tens of km~s$^{-1}$ \citep{1985ApJ...289..414F,2006ApJ...642L.169M}. Such a scenario has also been shown to hold not only for bigger flares, but also for microflare type outbursts \citep{2009ApJ...692..492B}. However, certain features in our events appear to be inconsistent with an evaporation scenario. For example, the speed of the `incoming' brightening in Event-1 is only $~$20~km~s$^{-1}$ whereas the speed for the 'reflected' one it is $~$50~ km~s$^{-1}$. Furthermore, in Event-3, we notice a delay of 20~s between the arrival of the incoming brightening and the onset of  the`reflected' one. These inconsistencies may also mean that these reflected brightenings are rather generated by new reconnection events occurring near those footpoints. Again, one would need co-spatial spectroscopic data to substantiate this. It is noteworthy that previous such reflection like phenomena seen in large coronal loops in AIA images, were interpreted as signatures of slow MHD waves \citep{2013ApJ...779L...7K,2016ApJ...828...72M}. One would need further observations to investigate this aspect.

Lastly, we address question how common propagating brightening associated with campfires are. We recall that by analyzing the same EUI dataset as in this work, \citet{2021arXiv210403382B} reported $\approx$150 campfires\footnote{Not all of these are unique events though, as any given campfire could be detected multiple times during its lifetime.} which are more than 5 EUI pixels in size (0.2 Mm$^{2}$) and live more than 5 EUI frames (25 s). At the same time, by visual inspections we could only find six events where prominent and systematic propagation signatures are present. It must be emphasized here that apart from the six cases that we have presented in this work, there are a few more events in our data where we can visually recognize the propagation signatures, but their signals are too ambiguous to establish anything quantitatively. Thus, at this moment, we are inclined to conclude that these propagating type localized brightenings are probably not a common phenomenon in quiet-Sun corona. However, with the availability of data with better spatial (and temporal) resolution from EUI one may possibly be able to resolve the above disparity. Such data will be available once Solar Orbiter enters its nominal mission phase. Furthermore, it will get as close as 0.3 AU to the Sun during perihelia, for the first time in early 2022 and the spatial resolution will improve further. 

\section{Acknowledgments}
We thank the anonymous referee for helpful comments on the manuscript. This project has received funding from the European Research Council (ERC) under the European Union’s Horizon 2020 research and innovation programme (grant agreement No 695075). Solar Orbiter is a mission of international cooperation between ESA and NASA, operated by ESA. The EUI instrument was built by CSL, IAS, MPS, MSSL/UCL, PMOD/WRC, ROB, LCF/IO with funding from the Belgian Federal Science Policy Office (BELSPO/PRODEX PEA 4000112292); the Centre National d’Etudes Spatiales (CNES); the UK Space Agency (UKSA); the Bundesministerium f\"{u}r Wirtschaft und Energie (BMWi) through the Deutsches Zentrum f\"{u}r Luft- und Raumfahrt (DLR); and the Swiss Space Office (SSO).

 \bibliographystyle{aa}
\bibliography{ref_int_prop_campfire}

\begin{thebibliography}{43}
\expandafter\ifx\csname natexlab\endcsname\relax\def\natexlab#1{#1}\fi

\bibitem[{{Alpert} {et~al.}(2016){Alpert}, {Tiwari}, {Moore}, {Winebarger}, \&
  {Savage}}]{2016ApJ...822...35A}
{Alpert}, S.~E., {Tiwari}, S.~K., {Moore}, R.~L., {Winebarger}, A.~R., \&
  {Savage}, S.~L. 2016, \apj, 822, 35

\bibitem[{{Aschwanden} {et~al.}(2000{\natexlab{a}}){Aschwanden}, {Nightingale},
  {Tarbell}, \& {Wolfson}}]{2000ApJ...535.1027A}
{Aschwanden}, M.~J., {Nightingale}, R.~W., {Tarbell}, T.~D., \& {Wolfson},
  C.~J. 2000{\natexlab{a}}, \apj, 535, 1027

\bibitem[{{Aschwanden} {et~al.}(2000{\natexlab{b}}){Aschwanden}, {Tarbell},
  {Nightingale}, {Schrijver}, {Title}, {Kankelborg}, {Martens}, \&
  {Warren}}]{2000ApJ...535.1047A}
{Aschwanden}, M.~J., {Tarbell}, T.~D., {Nightingale}, R.~W., {et~al.}
  2000{\natexlab{b}}, \apj, 535, 1047

\bibitem[{{Berghmans} {et~al.}(2021){Berghmans}, {Auch{\`e}re}, {Long},
  {Soubri{\'e}}, {Zhukov}, {Mierla}, {Sch{\"u}hle}, {Antolin}, {Parenti},
  {Harra}, {Podladchikova}, {Aznar Cuadrado}, {Buchlin}, {Dolla}, {Verbeeck},
  {Gissot}, {Teriaca}, {Haberreiter}, {Katsiyannis}, {Rodriguez}, {Kraaikamp},
  {Smith}, {Stegen}, {Rochus}, {Halain}, {Jacques}, {Thompson}, \&
  {Inhester}}]{2021arXiv210403382B}
{Berghmans}, D., {Auch{\`e}re}, F., {Long}, D.~M., {et~al.} 2021, arXiv
  e-prints, arXiv:2104.03382

\bibitem[{{Berghmans} \& {Clette}(1999)}]{1999SoPh..186..207B}
{Berghmans}, D. \& {Clette}, F. 1999, \solphys, 186, 207

\bibitem[{{Berghmans} {et~al.}(1998){Berghmans}, {Clette}, \&
  {Moses}}]{1998A&A...336.1039B}
{Berghmans}, D., {Clette}, F., \& {Moses}, D. 1998, \aap, 336, 1039

\bibitem[{{Brosius} \& {Holman}(2009)}]{2009ApJ...692..492B}
{Brosius}, J.~W. \& {Holman}, G.~D. 2009, \apj, 692, 492

\bibitem[{{Brueckner} \& {Bartoe}(1983)}]{1983ApJ...272..329B}
{Brueckner}, G.~E. \& {Bartoe}, J. D.~F. 1983, \apj, 272, 329

\bibitem[{{Chen} {et~al.}(2021){Chen}, {Przybylski}, {Peter}, {Tian},
  {Auch{\`e}re}, \& {Berghmans}}]{2021arXiv210410940C}
{Chen}, Y., {Przybylski}, D., {Peter}, H., {et~al.} 2021, arXiv e-prints,
  arXiv:2104.10940

\bibitem[{{Chitta} {et~al.}(2020){Chitta}, {Peter}, {Priest}, \&
  {Solanki}}]{2020A&A...644A.130C}
{Chitta}, L.~P., {Peter}, H., {Priest}, E.~R., \& {Solanki}, S.~K. 2020, \aap,
  644, A130

\bibitem[{{Chitta} {et~al.}(2021{\natexlab{a}}){Chitta}, {Peter}, \&
  {Young}}]{2021A&A...647A.159C}
{Chitta}, L.~P., {Peter}, H., \& {Young}, P.~R. 2021{\natexlab{a}}, \aap, 647,
  A159

\bibitem[{{Chitta} {et~al.}(2017){Chitta}, {Peter}, {Young}, \&
  {Huang}}]{2017A&A...605A..49C}
{Chitta}, L.~P., {Peter}, H., {Young}, P.~R., \& {Huang}, Y.~M. 2017, \aap,
  605, A49

\bibitem[{{Chitta} {et~al.}(2021{\natexlab{b}}){Chitta}, {Solanki}, {Peter},
  {Aznar Cuadrado}, {Teriaca}, {Sch{\"u}hle}, {Auch{\`e}re}, {Berghmans},
  {Kraaikamp}, {Gissot}, \& {Verbeeck}}]{2021arXiv210915106C}
{Chitta}, L.~P., {Solanki}, S.~K., {Peter}, H., {et~al.} 2021{\natexlab{b}},
  arXiv e-prints, arXiv:2109.15106

\bibitem[{{De Pontieu} {et~al.}(2014){De Pontieu}, {Title}, {Lemen}, {Kushner},
  {Akin}, {Allard}, {Berger}, {Boerner}, {Cheung}, {Chou}, {Drake}, {Duncan},
  {Freeland}, {Heyman}, {Hoffman}, {Hurlburt}, {Lindgren}, {Mathur}, {Rehse},
  {Sabolish}, {Seguin}, {Schrijver}, {Tarbell}, {W{\"u}lser}, {Wolfson},
  {Yanari}, {Mudge}, {Nguyen-Phuc}, {Timmons}, {van Bezooijen}, {Weingrod},
  {Brookner}, {Butcher}, {Dougherty}, {Eder}, {Knagenhjelm}, {Larsen},
  {Mansir}, {Phan}, {Boyle}, {Cheimets}, {DeLuca}, {Golub}, {Gates}, {Hertz},
  {McKillop}, {Park}, {Perry}, {Podgorski}, {Reeves}, {Saar}, {Testa}, {Tian},
  {Weber}, {Dunn}, {Eccles}, {Jaeggli}, {Kankelborg}, {Mashburn}, {Pust},
  {Springer}, {Carvalho}, {Kleint}, {Marmie}, {Mazmanian}, {Pereira}, {Sawyer},
  {Strong}, {Worden}, {Carlsson}, {Hansteen}, {Leenaarts}, {Wiesmann},
  {Aloise}, {Chu}, {Bush}, {Scherrer}, {Brekke}, {Martinez-Sykora}, {Lites},
  {McIntosh}, {Uitenbroek}, {Okamoto}, {Gummin}, {Auker}, {Jerram}, {Pool}, \&
  {Waltham}}]{2014SoPh..289.2733D}
{De Pontieu}, B., {Title}, A.~M., {Lemen}, J.~R., {et~al.} 2014, \solphys, 289,
  2733

\bibitem[{{Dere} {et~al.}(1989){Dere}, {Bartoe}, \&
  {Brueckner}}]{1989SoPh..123...41D}
{Dere}, K.~P., {Bartoe}, J. D.~F., \& {Brueckner}, G.~E. 1989, \solphys, 123,
  41

\bibitem[{{Fisher} {et~al.}(1985){Fisher}, {Canfield}, \&
  {McClymont}}]{1985ApJ...289..414F}
{Fisher}, G.~H., {Canfield}, R.~C., \& {McClymont}, A.~N. 1985, \apj, 289, 414

\bibitem[{{Galsgaard} {et~al.}(2017){Galsgaard}, {Madjarska},
  {Moreno-Insertis}, {Huang}, \& {Wiegelmann}}]{2017A&A...606A..46G}
{Galsgaard}, K., {Madjarska}, M.~S., {Moreno-Insertis}, F., {Huang}, Z., \&
  {Wiegelmann}, T. 2017, \aap, 606, A46

\bibitem[{{Harrison}(1997)}]{1997SoPh..175..467H}
{Harrison}, R.~A. 1997, \solphys, 175, 467

\bibitem[{{Hudson}(1991)}]{1991SoPh..133..357H}
{Hudson}, H.~S. 1991, \solphys, 133, 357

\bibitem[{{Innes} {et~al.}(2015){Innes}, {Guo}, {Huang}, \&
  {Bhattacharjee}}]{2015ApJ...813...86I}
{Innes}, D.~E., {Guo}, L.~J., {Huang}, Y.~M., \& {Bhattacharjee}, A. 2015,
  \apj, 813, 86

\bibitem[{{Innes} {et~al.}(1997){Innes}, {Inhester}, {Axford}, \&
  {Wilhelm}}]{1997Natur.386..811I}
{Innes}, D.~E., {Inhester}, B., {Axford}, W.~I., \& {Wilhelm}, K. 1997, \nat,
  386, 811

\bibitem[{{Kobayashi} {et~al.}(2014){Kobayashi}, {Cirtain}, {Winebarger},
  {Korreck}, {Golub}, {Walsh}, {De Pontieu}, {DeForest}, {Title}, {Kuzin},
  {Savage}, {Beabout}, {Beabout}, {Podgorski}, {Caldwell}, {McCracken},
  {Ordway}, {Bergner}, {Gates}, {McKillop}, {Cheimets}, {Platt}, {Mitchell}, \&
  {Windt}}]{2014SoPh..289.4393K}
{Kobayashi}, K., {Cirtain}, J., {Winebarger}, A.~R., {et~al.} 2014, \solphys,
  289, 4393

\bibitem[{{Krucker} {et~al.}(1997){Krucker}, {Benz}, {Bastian}, \&
  {Acton}}]{1997ApJ...488..499K}
{Krucker}, S., {Benz}, A.~O., {Bastian}, T.~S., \& {Acton}, L.~W. 1997, \apj,
  488, 499

\bibitem[{{Kumar} {et~al.}(2013){Kumar}, {Innes}, \&
  {Inhester}}]{2013ApJ...779L...7K}
{Kumar}, P., {Innes}, D.~E., \& {Inhester}, B. 2013, \apjl, 779, L7

\bibitem[{{Lemen} {et~al.}(2012){Lemen}, {Title}, {Akin}, {Boerner}, {Chou},
  {Drake}, {Duncan}, {Edwards}, {Friedlaender}, {Heyman}, {Hurlburt}, {Katz},
  {Kushner}, {Levay}, {Lindgren}, {Mathur}, {McFeaters}, {Mitchell}, {Rehse},
  {Schrijver}, {Springer}, {Stern}, {Tarbell}, {Wuelser}, {Wolfson}, {Yanari},
  {Bookbinder}, {Cheimets}, {Caldwell}, {Deluca}, {Gates}, {Golub}, {Park},
  {Podgorski}, {Bush}, {Scherrer}, {Gummin}, {Smith}, {Auker}, {Jerram},
  {Pool}, {Soufli}, {Windt}, {Beardsley}, {Clapp}, {Lang}, \&
  {Waltham}}]{2012SoPh..275...17L}
{Lemen}, J.~R., {Title}, A.~M., {Akin}, D.~J., {et~al.} 2012, \solphys, 275, 17

\bibitem[{{Madjarska}(2019)}]{2019LRSP...16....2M}
{Madjarska}, M.~S. 2019, Living Reviews in Solar Physics, 16, 2

\bibitem[{{Madjarska} {et~al.}(2003){Madjarska}, {Doyle}, {Teriaca}, \&
  {Banerjee}}]{2003A&A...398..775M}
{Madjarska}, M.~S., {Doyle}, J.~G., {Teriaca}, L., \& {Banerjee}, D. 2003,
  \aap, 398, 775

\bibitem[{{Mandal} {et~al.}(2016){Mandal}, {Yuan}, {Fang}, {Banerjee}, {Pant},
  \& {Van Doorsselaere}}]{2016ApJ...828...72M}
{Mandal}, S., {Yuan}, D., {Fang}, X., {et~al.} 2016, \apj, 828, 72

\bibitem[{{Milligan} {et~al.}(2006){Milligan}, {Gallagher}, {Mathioudakis}, \&
  {Keenan}}]{2006ApJ...642L.169M}
{Milligan}, R.~O., {Gallagher}, P.~T., {Mathioudakis}, M., \& {Keenan}, F.~P.
  2006, \apjl, 642, L169

\bibitem[{{M{\"u}ller} {et~al.}(2020){M{\"u}ller}, {St. Cyr}, {Zouganelis},
  {Gilbert}, {Marsden}, {Nieves-Chinchilla}, {Antonucci}, {Auch{\`e}re},
  {Berghmans}, {Horbury}, {Howard}, {Krucker}, {Maksimovic}, {Owen}, {Rochus},
  {Rodriguez-Pacheco}, {Romoli}, {Solanki}, {Bruno}, {Carlsson}, {Fludra},
  {Harra}, {Hassler}, {Livi}, {Louarn}, {Peter}, {Sch{\"u}hle}, {Teriaca}, {del
  Toro Iniesta}, {Wimmer-Schweingruber}, {Marsch}, {Velli}, {De Groof},
  {Walsh}, \& {Williams}}]{2020A&A...642A...1M}
{M{\"u}ller}, D., {St. Cyr}, O.~C., {Zouganelis}, I., {et~al.} 2020, \aap, 642,
  A1

\bibitem[{{Panesar} {et~al.}(2021){Panesar}, {Tiwari}, {Berghmans}, {Cheung},
  {Muller}, {Auchere}, \& {Zhukov}}]{2021arXiv211006846P}
{Panesar}, N.~K., {Tiwari}, S.~K., {Berghmans}, D., {et~al.} 2021, arXiv
  e-prints, arXiv:2110.06846

\bibitem[{{Peter} {et~al.}(2014){Peter}, {Tian}, {Curdt}, {Schmit}, {Innes},
  {De Pontieu}, {Lemen}, {Title}, {Boerner}, {Hurlburt}, {Tarbell}, {Wuelser},
  {Mart{\'\i}nez-Sykora}, {Kleint}, {Golub}, {McKillop}, {Reeves}, {Saar},
  {Testa}, {Kankelborg}, {Jaeggli}, {Carlsson}, \&
  {Hansteen}}]{2014Sci...346C.315P}
{Peter}, H., {Tian}, H., {Curdt}, W., {et~al.} 2014, Science, 346, 1255726

\bibitem[{{Rachmeler} {et~al.}(2019){Rachmeler}, {Winebarger}, {Savage},
  {Golub}, {Kobayashi}, {Vigil}, {Brooks}, {Cirtain}, {De Pontieu}, {McKenzie},
  {Morton}, {Peter}, {Testa}, {Tiwari}, {Walsh}, {Warren}, {Alexander},
  {Ansell}, {Beabout}, {Beabout}, {Bethge}, {Champey}, {Cheimets}, {Cooper},
  {Creel}, {Gates}, {Gomez}, {Guillory}, {Haight}, {Hogue}, {Holloway}, {Hyde},
  {Kenyon}, {Marshall}, {McCracken}, {McCracken}, {Mitchell}, {Ordway}, {Owen},
  {Ranganathan}, {Robertson}, {Payne}, {Podgorski}, {Pryor}, {Samra}, {Sloan},
  {Soohoo}, {Steele}, {Thompson}, {Thornton}, {Watkinson}, \&
  {Windt}}]{2019SoPh..294..174R}
{Rachmeler}, L.~A., {Winebarger}, A.~R., {Savage}, S.~L., {et~al.} 2019,
  \solphys, 294, 174

\bibitem[{{Rochus} {et~al.}(2020){Rochus}, {Auch{\`e}re}, {Berghmans}, {Harra},
  {Schmutz}, {Sch{\"u}hle}, {Addison}, {Appourchaux}, {Aznar Cuadrado},
  {Baker}, {Barbay}, {Bates}, {BenMoussa}, {Bergmann}, {Beurthe}, {Borgo},
  {Bonte}, {Bouzit}, {Bradley}, {B{\"u}chel}, {Buchlin}, {B{\"u}chner},
  {Cab{\'e}}, {Cadiergues}, {Chaigneau}, {Chares}, {Choque Cortez}, {Coker},
  {Condamin}, {Coumar}, {Curdt}, {Cutler}, {Davies}, {Davison}, {Defise}, {Del
  Zanna}, {Delmotte}, {Delouille}, {Dolla}, {Dumesnil}, {D{\"u}rig}, {Enge},
  {Fran{\c{c}}ois}, {Fourmond}, {Gillis}, {Giordanengo}, {Gissot}, {Green},
  {Guerreiro}, {Guilbaud}, {Gyo}, {Haberreiter}, {Hafiz}, {Hailey}, {Halain},
  {Hansotte}, {Hecquet}, {Heerlein}, {Hellin}, {Hemsley}, {Hermans}, {Hervier},
  {Hochedez}, {Houbrechts}, {Ihsan}, {Jacques}, {J{\'e}r{\^o}me}, {Jones},
  {Kahle}, {Kennedy}, {Klaproth}, {Kolleck}, {Koller}, {Kotsialos},
  {Kraaikamp}, {Langer}, {Lawrenson}, {Le Clech'}, {Lenaerts}, {Liebecq},
  {Linder}, {Long}, {Mampaey}, {Markiewicz-Innes}, {Marquet}, {Marsch},
  {Matthews}, {Mazy}, {Mazzoli}, {Meining}, {Meltchakov}, {Mercier}, {Meyer},
  {Monecke}, {Monfort}, {Morinaud}, {Moron}, {Mountney}, {M{\"u}ller},
  {Nicula}, {Parenti}, {Peter}, {Pfiffner}, {Philippon}, {Phillips},
  {Plesseria}, {Pylyser}, {Rabecki}, {Ravet-Krill}, {Rebellato}, {Renotte},
  {Rodriguez}, {Roose}, {Rosin}, {Rossi}, {Roth}, {Rouesnel}, {Roulliay},
  {Rousseau}, {Ruane}, {Scanlan}, {Schlatter}, {Seaton}, {Silliman}, {Smit},
  {Smith}, {Solanki}, {Spescha}, {Spencer}, {Stegen}, {Stockman}, {Szwec},
  {Tamiatto}, {Tandy}, {Teriaca}, {Theobald}, {Tychon}, {van Driel-Gesztelyi},
  {Verbeeck}, {Vial}, {Werner}, {West}, {Westwood}, {Wiegelmann}, {Willis},
  {Winter}, {Zerr}, {Zhang}, \& {Zhukov}}]{2020A&A...642A...8R}
{Rochus}, P., {Auch{\`e}re}, F., {Berghmans}, D., {et~al.} 2020, \aap, 642, A8

\bibitem[{{Solanki} {et~al.}(2020){Solanki}, {del Toro Iniesta}, {Woch},
  {Gandorfer}, {Hirzberger}, {Alvarez-Herrero}, {Appourchaux}, {Mart{\'\i}nez
  Pillet}, {P{\'e}rez-Grande}, {Sanchis Kilders}, {Schmidt}, {G{\'o}mez Cama},
  {Michalik}, {Deutsch}, {Fernandez-Rico}, {Grauf}, {Gizon}, {Heerlein},
  {Kolleck}, {Lagg}, {Meller}, {M{\"u}ller}, {Sch{\"u}hle}, {Staub}, {Albert},
  {Alvarez Copano}, {Beckmann}, {Bischoff}, {Busse}, {Enge}, {Frahm},
  {Germerott}, {Guerrero}, {L{\"o}ptien}, {Meierdierks}, {Oberdorfer},
  {Papagiannaki}, {Ramanath}, {Schou}, {Werner}, {Yang}, {Zerr}, {Bergmann},
  {Bochmann}, {Heinrichs}, {Meyer}, {Monecke}, {M{\"u}ller}, {Sperling},
  {{\'A}lvarez Garc{\'\i}a}, {Aparicio}, {Balaguer Jim{\'e}nez}, {Bellot
  Rubio}, {Cobos Carracosa}, {Girela}, {Hern{\'a}ndez Exp{\'o}sito}, {Herranz},
  {Labrousse}, {L{\'o}pez Jim{\'e}nez}, {Orozco Su{\'a}rez}, {Ramos},
  {Barandiar{\'a}n}, {Bastide}, {Campuzano}, {Cebollero}, {D{\'a}vila},
  {Fern{\'a}ndez-Medina}, {Garc{\'\i}a Parejo}, {Garranzo-Garc{\'\i}a},
  {Laguna}, {Mart{\'\i}n}, {Navarro}, {N{\'u}{\~n}ez Peral}, {Royo},
  {S{\'a}nchez}, {Silva-L{\'o}pez}, {Vera}, {Villanueva}, {Fourmond}, {de
  Galarreta}, {Bouzit}, {Hervier}, {Le Clec'h}, {Szwec}, {Chaigneau},
  {Buttice}, {Dominguez-Tagle}, {Philippon}, {Boumier}, {Le Cocguen},
  {Baranjuk}, {Bell}, {Berkefeld}, {Baumgartner}, {Heidecke}, {Maue}, {Nakai},
  {Scheiffelen}, {Sigwarth}, {Soltau}, {Volkmer}, {Blanco Rodr{\'\i}guez},
  {Domingo}, {Ferreres Sabater}, {Gasent Blesa}, {Rodr{\'\i}guez
  Mart{\'\i}nez}, {Osorno Caudel}, {Bosch}, {Casas}, {Carmona}, {Herms},
  {Roma}, {Alonso}, {G{\'o}mez-Sanjuan}, {Piqueras}, {Torralbo}, {Fiethe},
  {Guan}, {Lange}, {Michel}, {Bonet}, {Fahmy}, {M{\"u}ller}, \&
  {Zouganelis}}]{2020A&A...642A..11S}
{Solanki}, S.~K., {del Toro Iniesta}, J.~C., {Woch}, J., {et~al.} 2020, \aap,
  642, A11

\bibitem[{{Spice Consortium} {et~al.}(2020){Spice Consortium}, {Anderson},
  {Appourchaux}, {Auch{\`e}re}, {Aznar Cuadrado}, {Barbay}, {Baudin},
  {Beardsley}, {Bocchialini}, {Borgo}, {Bruzzi}, {Buchlin}, {Burton},
  {B{\"u}chel}, {Caldwell}, {Caminade}, {Carlsson}, {Curdt}, {Davenne},
  {Davila}, {Deforest}, {Del Zanna}, {Drummond}, {Dubau}, {Dumesnil}, {Dunn},
  {Eccleston}, {Fludra}, {Fredvik}, {Gabriel}, {Giunta}, {Gottwald}, {Griffin},
  {Grundy}, {Guest}, {Gyo}, {Haberreiter}, {Hansteen}, {Harrison}, {Hassler},
  {Haugan}, {Howe}, {Janvier}, {Klein}, {Koller}, {Kucera}, {Kouliche},
  {Marsch}, {Marshall}, {Marshall}, {Matthews}, {McQuirk}, {Meining},
  {Mercier}, {Morris}, {Morse}, {Munro}, {Parenti}, {Pastor-Santos}, {Peter},
  {Pfiffner}, {Phelan}, {Philippon}, {Richards}, {Rogers}, {Sawyer},
  {Schlatter}, {Schmutz}, {Sch{\"u}hle}, {Shaughnessy}, {Sidher}, {Solanki},
  {Speight}, {Spescha}, {Szwec}, {Tamiatto}, {Teriaca}, {Thompson}, {Tosh},
  {Tustain}, {Vial}, {Walls}, {Waltham}, {Wimmer-Schweingruber}, {Woodward},
  {Young}, {de Groof}, {Pacros}, {Williams}, \&
  {M{\"u}ller}}]{2020A&A...642A..14S}
{Spice Consortium}, {Anderson}, M., {Appourchaux}, T., {et~al.} 2020, \aap,
  642, A14

\bibitem[{{Tian} {et~al.}(2014){Tian}, {Kleint}, {Peter}, {Weber}, {Testa},
  {DeLuca}, {Golub}, \& {Schanche}}]{2014ApJ...790L..29T}
{Tian}, H., {Kleint}, L., {Peter}, H., {et~al.} 2014, \apjl, 790, L29

\bibitem[{{Tiwari} {et~al.}(2019){Tiwari}, {Panesar}, {Moore}, {De Pontieu},
  {Winebarger}, {Golub}, {Savage}, {Rachmeler}, {Kobayashi}, {Testa}, {Warren},
  {Brooks}, {Cirtain}, {McKenzie}, {Morton}, {Peter}, \&
  {Walsh}}]{2019ApJ...887...56T}
{Tiwari}, S.~K., {Panesar}, N.~K., {Moore}, R.~L., {et~al.} 2019, \apj, 887, 56

\bibitem[{{Winebarger} {et~al.}(2013){Winebarger}, {Walsh}, {Moore}, {De
  Pontieu}, {Hansteen}, {Cirtain}, {Golub}, {Kobayashi}, {Korreck}, {DeForest},
  {Weber}, {Title}, \& {Kuzin}}]{2013ApJ...771...21W}
{Winebarger}, A.~R., {Walsh}, R.~W., {Moore}, R., {et~al.} 2013, \apj, 771, 21

\bibitem[{{Young} {et~al.}(2018){Young}, {Tian}, {Peter}, {Rutten}, {Nelson},
  {Huang}, {Schmieder}, {Vissers}, {Toriumi}, {Rouppe van der Voort},
  {Madjarska}, {Danilovic}, {Berlicki}, {Chitta}, {Cheung}, {Madsen},
  {Reardon}, {Katsukawa}, \& {Heinzel}}]{2018SSRv..214..120Y}
{Young}, P.~R., {Tian}, H., {Peter}, H., {et~al.} 2018, \ssr, 214, 120

\bibitem[{{Zacharias} {et~al.}(2011){Zacharias}, {Peter}, \&
  {Bingert}}]{2011A&A...532A.112Z}
{Zacharias}, P., {Peter}, H., \& {Bingert}, S. 2011, \aap, 532, A112

\bibitem[{{Zhang} {et~al.}(2012){Zhang}, {Chen}, {Guo}, {Fang}, \&
  {Ding}}]{2012ApJ...746...19Z}
{Zhang}, Q.~M., {Chen}, P.~F., {Guo}, Y., {Fang}, C., \& {Ding}, M.~D. 2012,
  \apj, 746, 19

\bibitem[{{Zhukov} {et~al.}(2021){Zhukov}, {Mierla}, {Auch{\`e}re}, {Gissot},
  {Rodriguez}, {Soubri{\'e}}, {Thompson}, {Inhester}, {Nicula}, {Antolin},
  {Parenti}, {Buchlin}, {Barczynski}, {Verbeeck}, {Kraaikamp}, {Smith},
  {Stegen}, {Dolla}, {Harra}, {Long}, {Sch{\"u}hle}, {Podladchikova}, {Aznar
  Cuadrado}, {Teriaca}, {Haberreiter}, {Katsiyannis}, {Rochus}, {Halain},
  {Jacques}, \& {Berghmans}}]{2021arXiv210902169Z}
{Zhukov}, A.~N., {Mierla}, M., {Auch{\`e}re}, F., {et~al.} 2021, arXiv
  e-prints, arXiv:2109.02169

\end{thebibliography}

 \begin{appendix}
 
\section{Two more cases: Event-5 and 6}\label{S:appendix}

\begin{figure}[!htb]
\centering
\includegraphics[width=0.45\textwidth,clip,trim=0cm 0cm 0cm 0cm]{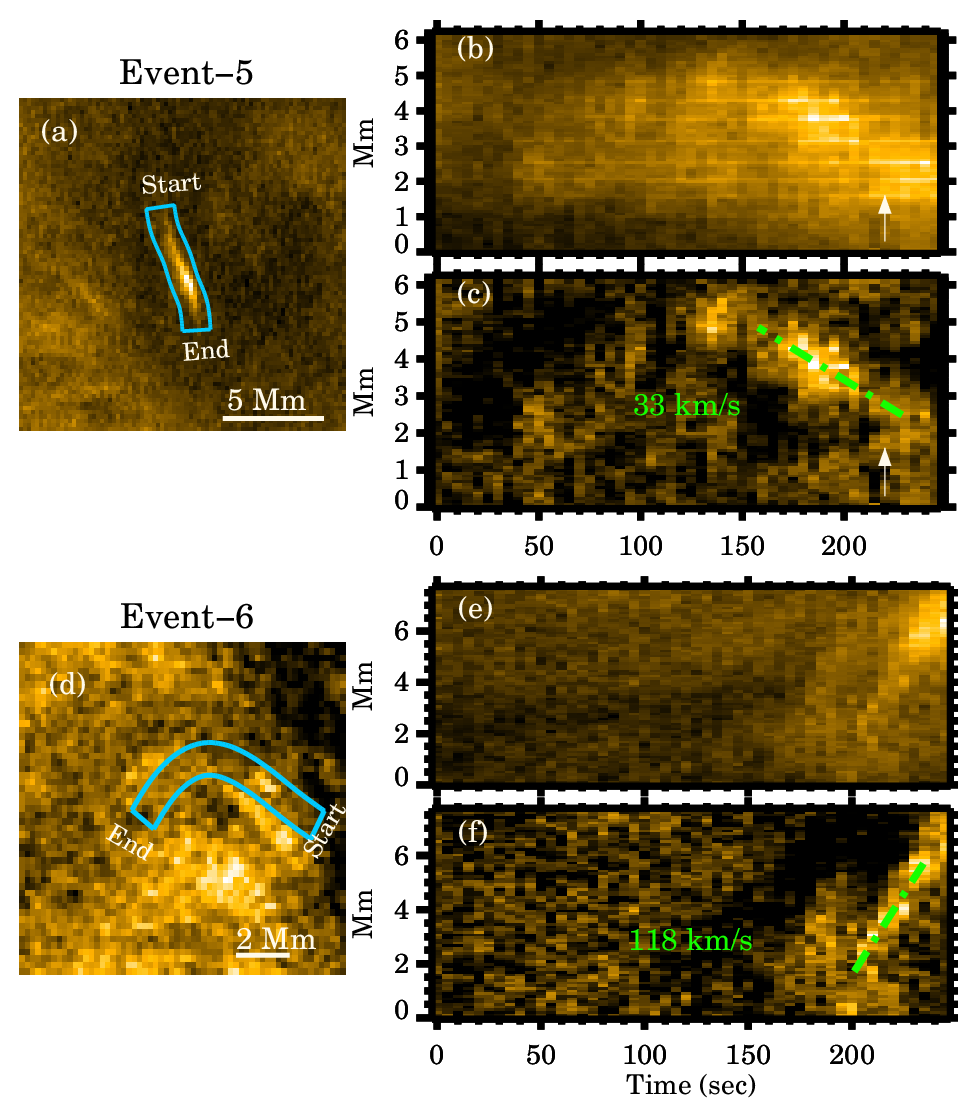}
\caption{Overview of Event-5 (in {\it panels-a,c}) and Event-6 (in {\it panels-d,f}). The layout for each event is the same as in Fig.~\ref{fig2}a-c. The arrows in panels b and c point towards a specific ridge in these maps as discussed in appendix~\ref{S:appendix}.}
\label{fig1_app}

\end{figure}

Propagation characteristics that we find in Event-5 (Fig.~\ref{fig1_app}a) are generally similar to that of the Event-3, but with one significant exception. In this case, a small bright patch (of a few pixels) appears to glide over another comparatively larger (extended) bright feature. In the space-time map (Fig~\ref{fig1_app}b), this appears as a bright arc-like structure riding on top of a diffuse bright background. A closer look at this X-T map reveals the fragmentary nature of the  ridge propagating from "Start" to "End" (i.e. from 0 to 5 Mm). This is due to the fact that during this outward motion, the small bright patch keeps appearing and disappearing multiple times (see also the event movie available online). On the other hand, the return propagation ridge is prominent and continuous. The propagation speed is measured to be 33 km~s$^{-1}$ (Fig~\ref{fig1_app}c). Additionally, we also notice a small upward moving ridge near $t$=225~s (highlighted with the arrows in Fig~\ref{fig1_app}b,c). This occurs due to a new bright feature that appears near what we assume to be the `end' footpoint and moves upwards to meet the return brightening at $t$=245~s.

 Event-6 (Fig.~\ref{fig1_app}d) can be described as an event similar to Event-4, but with some differences. In this event, we find a small but isolated bright blob-like feature (0.7 Mm$^{2}$ in size) propagating from one footpoint to the other, following an arched path. We outline this propagation path by the curved slit shown in Fig.~\ref{fig1_app}d. The appearance of this blob-like feature is somewhat similar to the dot-like brightenings reported in \citet{2019ApJ...887...56T}. Such a blob also has been found in a 3D numerical model, albeit in an active region setting \citep{2011A&A...532A.112Z}.
 The bright blob appears close to the end of this observation (starts at $t$=200~s) and continues to move till the very last frame where it is seen to merge with an adjacent bright feature. Interestingly, this blob propagates with a significantly larger speed of 118~km~s$^{-1}$) (Fig.~\ref{fig1_app}f), which is comparable to the local sound speed (assuming again the plasma to be at a million degrees Kelvin). The speed of this brightening is comparable to the plasma jet speeds reported in \citet{2021arXiv210915106C}.

\section{AIA view of propagating campfires}
 Considering the similarities between the AIA 171~{\AA} and \HRIEUV 174~{\AA} channels \cite[both these passbands capture similar plasma temperatures;][their Fig.\,1]{2021arXiv210410940C}, it is natural to look for the signatures of these propagating campfires also in AIA 171~{\AA} data. At this point, we recall that the AIA spatial resolution is only about 900 km as compared to 400 km in {\HRIEUV} for the observations analysed here. Furthermore, the AIA EUV image cadence (of 12 s) is more than twice of that of this \HRIEUV campaign (5 s). Therefore, these propagating campfires which typically have widths between 2-6 EUI pixels (and a characteristic propagation timescale of $\approx$40 s), are most likely to appear fuzzy and stationary in AIA 171~{\AA} data.
\begin{figure}[!htb]
\centering
\includegraphics[width=0.45\textwidth,clip,trim=0cm 0cm 0cm 0cm]{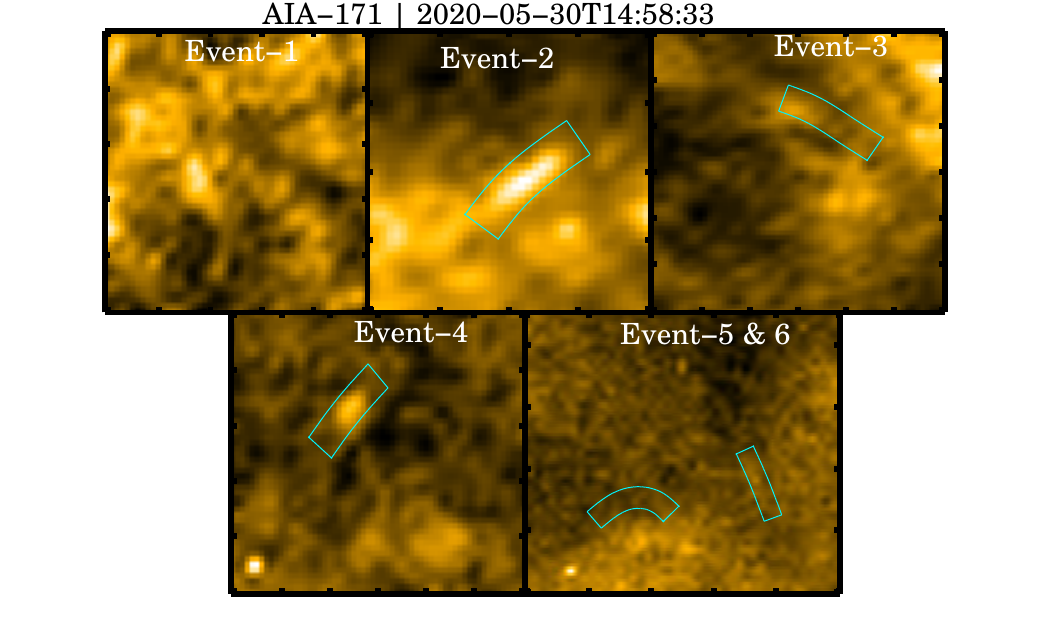}
\includegraphics[width=0.45\textwidth,clip,trim=0cm 0cm 0cm 0cm]{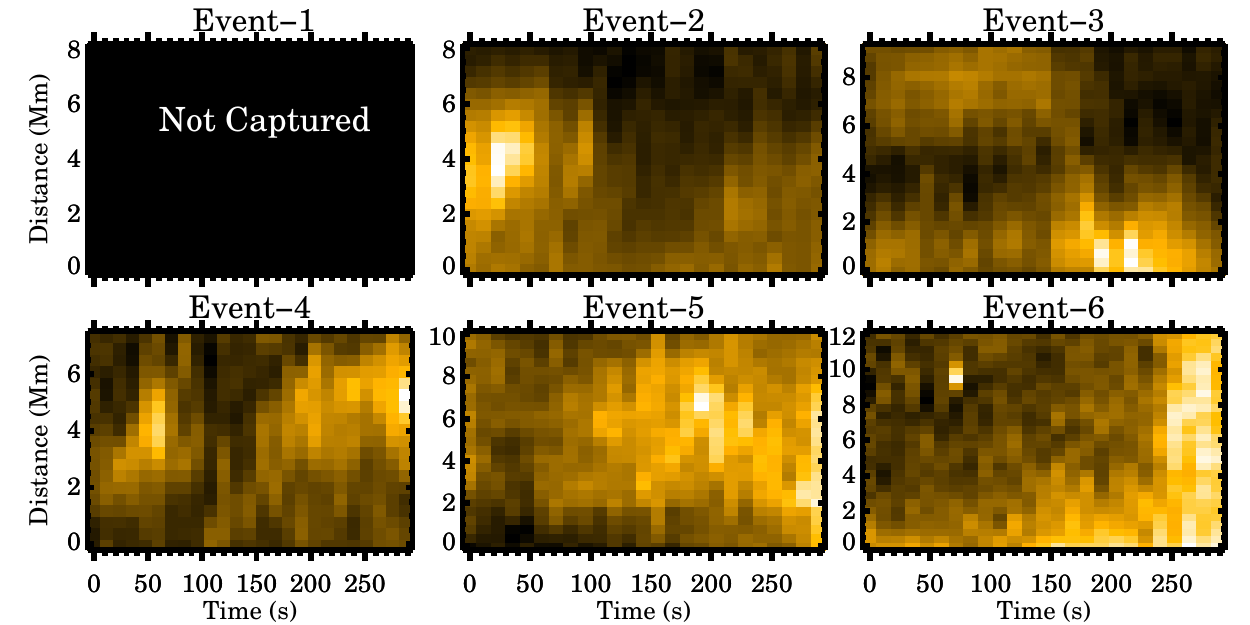}
\caption{AIA view of propagating campfires. The set of five panels at the top presents AIA snapshots of each of our six events. The curved
boxes shown in each of these panels mark the locations of the artificial slits used to derive temporal evolution in the X-T maps. These X-T maps are shown in the bottom set of panels. Because we could not identify Event-1 in AIA, we leave the corresponding X-T map blank. An animated version of this figure is available online.} 
\label{fig2_app}

\end{figure}
 As described earlier (in Sect.~\ref{sec_data}), at the time of this observation (i.e., on 30-May-2020) Solar Orbiter was located at a distance of 0.556 AU with an angle of 31.5$^o$ west in solar longitude from the Earth-Sun line. Thus, the same FOV was also co-observed by AIA. We again refer to the \citet{2021arXiv210403382B} paper for further details on this. In Fig.~\ref{fig2_app} we show the AIA snapshots of each of our EUI events and subsequently show corresponding X-T maps which are derived using the co-temporal and co-aligned AIA data. As can be seen from these AIA maps, propagating signatures are largely indistinguishable from the background and in-fact, some ridges can only be identified in hindsight i.e., only after seeing the EUI maps. In case of Event-1, we could only see the brightening in two AIA frames and as a result, could not trace out its propagation path. Therefore, from this analysis, we conclude that due to the resolution differences (both spatial and temporal), signatures of propagating campfires can not be identified unambiguously by only using AIA~171{\AA} image sequences.

\section{Signal-to-noise ratio}

\begin{figure}[!htb]
\centering
\includegraphics[width=0.42\textwidth,clip,trim=0cm 0cm 0cm 0cm]{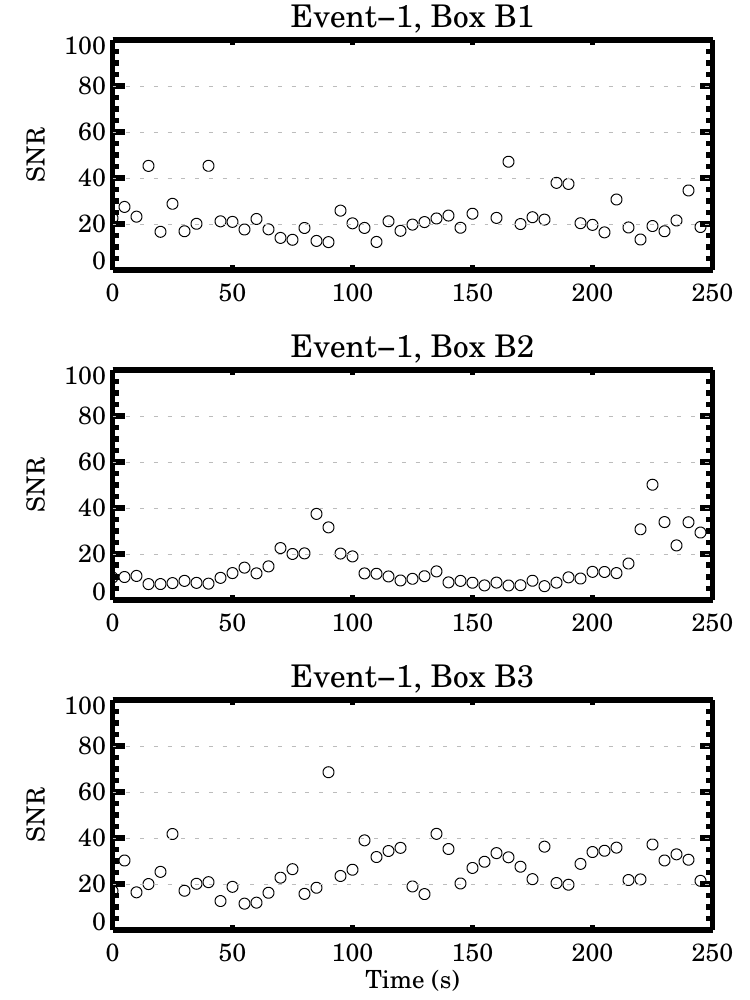}
\caption{Derived SNR values for Event-1. Locations of Box B1 (top panel), B2 (middle panel) and B3 (bottom panel) are the same as shown in Fig.~\ref{fig2}a.} 
\label{fig4_app}
\end{figure}

Although the spatial and temporal scales associated with these propagating brightenings are quite close to the instrumental limits, the {\HRIEUV} images have sufficient signal to noise to facilitate unambiguous detection of these events. To get a first order estimate of the noise level in each image, we calculate the mean intensity and the sigma (standard deviation) within a box that encompasses the campfire location. The signal-to-noise ratio (SNR) is then defined as the ratio between mean and sigma, i.e., SNR=mean/sigma. In Fig.~\ref{fig4_app}, we show the estimated values of SNR for Event1. These values of SNR underline our conclusion that the derived features are real signals.

\section{Light curves of remaining EUI events}

\begin{figure*}[!htb]
\centering
\includegraphics[width=0.85\textwidth,clip,trim=0cm 0.8cm 0cm 1cm]{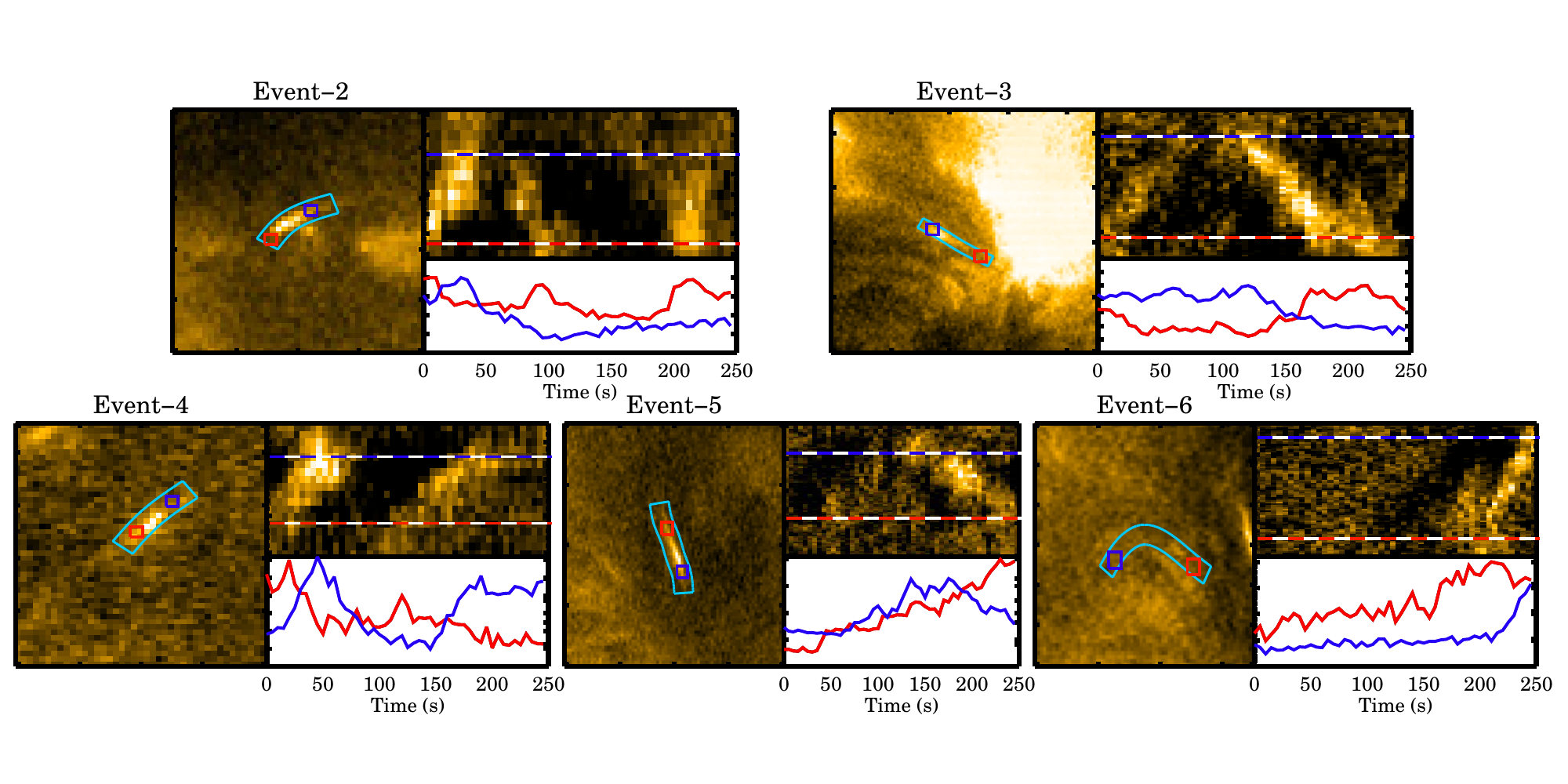}
\caption{EUI light curves of Event-2 to Event-6. In each of these panels, the curved box (in cyan) highlights the artificial slit wherein the blue and red boxes mark the locations from where the light curves are derived. The centers of these boxes are also overplotted on top of the associated X-T map as straight horizontal lines. On the bottom of the X-T map, the light curves are shown in red and blue colors.} 
\label{fig3_app}
\end{figure*}

Previously, in Sect.~\ref{sec_ev1} we have shown the light curves of Event-1 (Fig.~\ref{fig2}). In Fig.~\ref{fig3_app}, we now present the same for the remaining five cases (Events-2 to 6). In all of these events, the brightening was either present before the start of the observation (i.e., on the first frame) or the feature was still evolving on the last frame. So, the complete evolution could not be captured and hence, no new information is available through these light curves relative to what is already there in their corresponding X-T maps. 

 \end{appendix}


\end{document}